%% file: template.tex
\documentclass[preprint]{vgtc}                     

\onlineid{7121}

\vgtccategory{Research}
\preprinttext{Accepted for publication at IEEE VIS 2026 Workshop on GenAI, Agents, and the Future of VIS.}
\title{Semantic Action Graph: A Shared Representation for Agent Grounding and Human Interpretation of Sports Highlights}

\author{
\authororcid{Tica Lin}{0000-0002-2860-0871}\thanks{T. Lin, D. Chandran, G. Jagatap, A. Fanelli, D. Gunawan, and J. Kimball are with Dolby Laboratories. E-mail: tica.lin@dolby.com}, %
\authororcid{Deepak Chandran}{0009-0007-5306-9625},
\authororcid{ Gauri Jagatap}{0000-0001-7499-2581}, %
\authororcid{ Chen Chen}{0000-0003-3171-0657}\thanks{C. Chen is now with XPENG.}, %
\authororcid{Andrea Fanelli}{0009-0008-4349-2371}, %
\\ \authororcid{David Gunawan}{0009-0004-8606-3842}, %
and \authororcid{Josh Kimball}{0009-0009-9586-2044} %
\\[1ex]
Dolby Laboratories
\vspace{-6mm}
}

\authorfooter{
\item All authors are with Dolby Laboratories. *Email: tica.lin@dolby.com
}

\input{docs/0_abstract}


\teaser{
  \centering
\includegraphics[width=\linewidth]{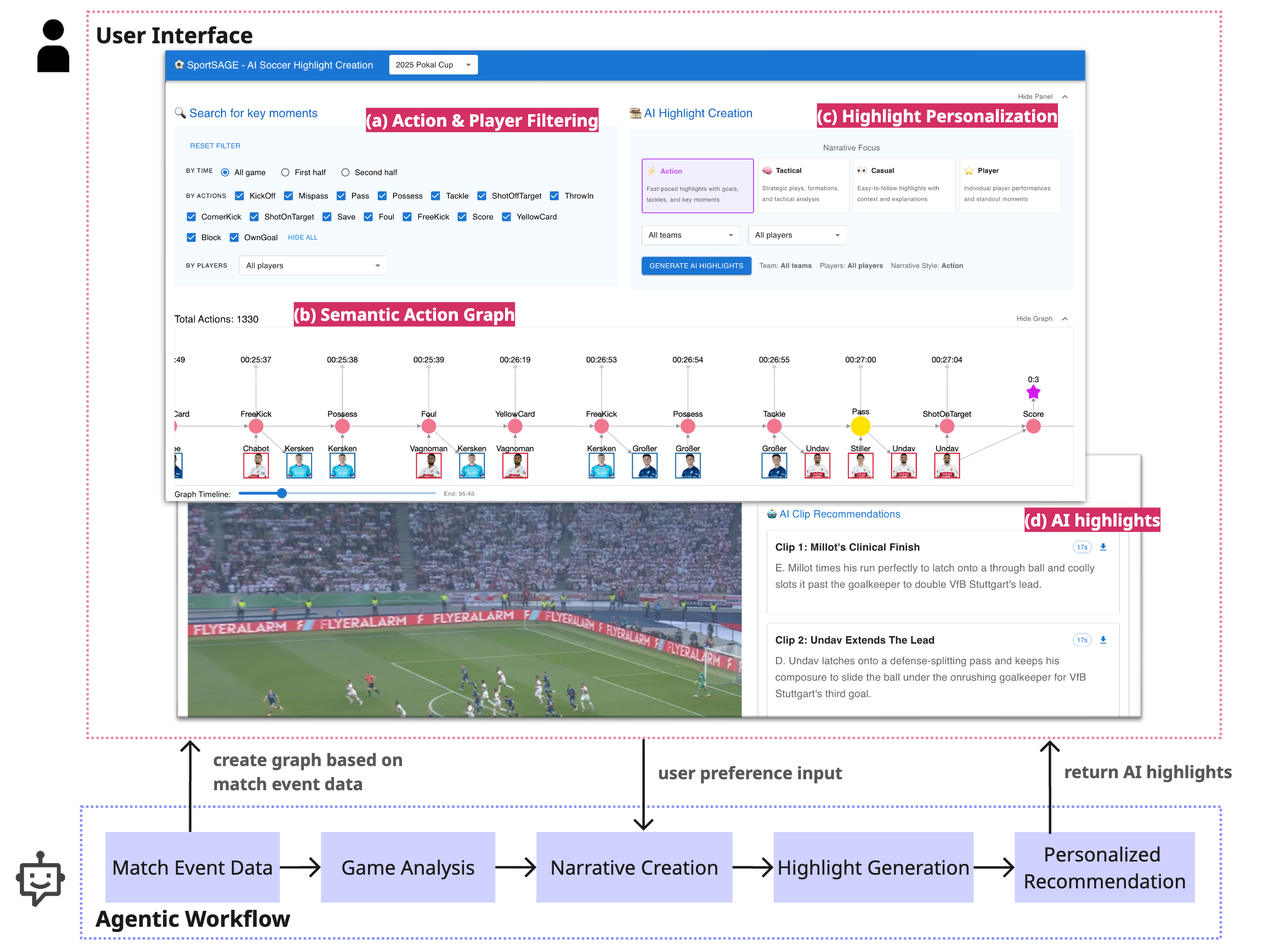}
  \caption{
   We design the semantic action graph as a shared representation for sports highlights, expressing a match as performer, action, recipient, moment, and state nodes joined by semantic edges. SportSAGE instantiates it with a highlight agent (bottom) to compute statistics, compose narratives, and return clips as node sequences with frame boundaries read from the graph. The same structure drives the interface (top) for  filtering (a) and navigating (b) the graph. Viewers can specify preferred narrative focus (c) and view the resulting personalized highlights  (d).
  }
  \label{fig:teaser}
}

\graphicspath{{figs/}{figures/}{pictures/}{images/}{./}} 

\usepackage{tabu}                      
\usepackage{booktabs}                  
\usepackage{lipsum}                    
\usepackage{mwe}                       
\usepackage{ccicons}                   

\usepackage{mathptmx}                  

\usepackage{graphicx}
\usepackage{tablefootnote}
\usepackage{comment}
\usepackage{mathptmx}                  

\usepackage{tikz}
\definecolor{buttonblue}{HTML}{3266C4}

\usepackage{enumitem}
\setlist[enumerate]{itemsep=0pt, topsep=0pt, partopsep=0pt, parsep=0pt}
\usepackage{enumitem}
\usepackage{times}     
\usepackage{xcolor}
 
\newcommand{\graph}{SportSAGE}
\usepackage{graphicx}
\newcommand{\Action}[1][]{\includegraphics[height=0.8em,#1]{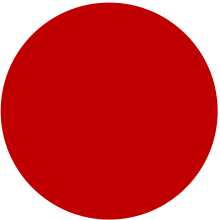}}
\newcommand{\Identity}[1][]{\includegraphics[height=0.8em,#1]{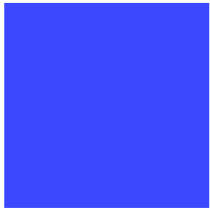}}
\newcommand{\Moment}[1][]{\includegraphics[height=0.8em,#1]{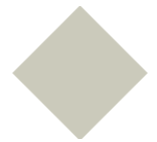}}

\newcommand{\Recipient}[1][]{\includegraphics[height=0.8em,#1]{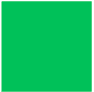}}
\newcommand{\State}[1][]{\includegraphics[height=0.8em,#1]{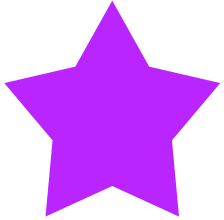}}

\begin{document}


\firstsection{Introduction}

\maketitle

\input{docs/1_intro}

\input{docs/2_related_work}
\input{docs/3_formative_study}

\input{docs/4_design}
\input{docs/5_system}
\input{docs/6_evaluation}

\input{docs/7_discussion}

\acknowledgments{
We acknowledge the Deutsche Fußball Liga (DFL) for granting access to official match data and video footage used in this research. We also thank the interviewees in our formative study for sharing their experiences.
}

\bibliographystyle{abbrv-doi-hyperref}

\bibliography{template}

\end{document}

%% file: docs/0_abstract.tex
\abstract{
Generative agents are increasingly used to select and narrate video highlights, but they typically operate over unstructured or frame-level representations. Their output is consequently difficult for a viewer to verify and steer toward individual preferences. We present the \emph{semantic action graph}, a lightweight domain schema that represents a sports match as performer, action, recipient, moment, and state nodes connected by role, temporal, and outcome edges. The schema demonstrates three key properties: 1) connected event sequences, 2) a shared, closed vocabulary, and 3) frame-addressable moments, making it suitable to serve two consumers at once: an agentic pipeline that composes narrated highlights, and a visual interface through which viewers query and inspect the same structure. We instantiate it in SportSAGE, a design probe pairing a four-module highlight pipeline with a graph interface, and report feedback from 12 soccer fans. Participants were satisfied with the quality of the generated highlights and narratives, and used the graph interface to search, navigate, and interpret the match highlights. These results provide early evidence that one small, human-readable schema can ground agent generation and support human interpretation at the same time.
}

%% file: docs/1_intro.tex
Sports highlights are central to how fans engage with a game beyond the live match, condensing key moments into a reel with narrative that explains their significance. Yet fans differ in what they want to see and how they want it framed—a single manually edited reel can't serve everyone. As a result, current highlights often miss the moments individual fans care about, offer little narrative context, and provide no effective way to find a specific play.

%
Recent systems use large language and vision-language models to automatically generate highlights and narratives. One line of work  selects or ranks the moments worth showing, by scoring frame importance~\cite{lee2025llmvs}, combining domain metrics with contextual reasoning~\cite{kang2025diamond}, or assisting an editor with language-based retrieval~\cite{wang2024lave}. Another line of work generates language to accompany footage already chosen, such as live commentary~\cite{andrews2024aicommentator} or answers to questions about a play~\cite{lee2025sportify,rao2025socceragent}.
These two halves are usually developed in isolation, and systems that produce both together are limited~\cite{barua2025lotus}.

Because the generated prose isn't linked to the underlying events, fans can't verify its accuracy or customize results meaningfully. If the narrative mentions a defense-splitting pass, nothing in the interface confirms it happened, who made it, or when. A fan wanting an earlier starting point or the passes leading to shots rather than the shots themselves, has no way to specify this, since the system doesn't expose those events. Existing pipelines work over frame embeddings and text tokens, which enable generation but aren't units a person can inspect or reference.

We attribute both limitations to the lack of a shared intermediate representation. 
Visualization tools have produced structured video indices that support browsing~\cite{matejka2014videolens,deng2021eventanchor}, but generative models don't consume them. 
Agentic pipelines that build explicit structures keep them internal, discarding them before output reaches the viewer~\cite{yang2023pvsg,huang2025mindpalace}. A representation that simultaneously grounds the AI pipeline and supports human interpretation remains absent.

To bridge the gap, we present the \emph{semantic action graph}, a lightweight schema that represents a sports match as performer, action, recipient, moment, and state nodes connected by role, temporal, and outcome relations.
Designed for both agent and human consumers, the graph captures connected event sequences using a closed vocabulary and frame-addressable moments. 
An agentic pipeline can use it to compute match statistics, compose narratives, and retrieve clips, while a visual interface renders the identical structure for viewers to filter, inspect, and navigate.
We demonstrate the schema through SportSAGE, a design probe for soccer highlights that pairs a four-module agentic pipeline with a graph-based interface. 
Grounded in three design goals derived from a formative study with six fans, SportSAGE takes structured match event data from an official league feed~\cite{DFL_OfficialMatchData} and produces personalized narrated highlights that viewers can verify and navigate.

We collected feedback from 12 soccer fans using SportSAGE on a professional match. Participants valued receiving customized highlights with varying narrative focuses of the same match, and used the graph layer to inspect generated clips and check the narratives against the video. These results serve as early evidence that the schema can ground agentic highlight creation and guide human interaction with its output.

This work contributes the following: (1) the semantic action graph, a domain schema that serve both an agentic pipeline and a viewer-facing interface; (2) SportSAGE, a design probe demonstrating the schema in soccer highlight generation; and (3) user feedback on narrative personalization and the value of a graph layer for interpreting and interacting with generated video content.

%% file: docs/2_related_work.tex
\section{Related Work}

\subsection{Sports highlight generation}

Automatically extracting highlights from sports video has been studied for over two decades, through proxies for excitement such as commentator speech and crowd noise~\cite{rui2000baseball}, modeled viewer arousal~\cite{hanjalic2005excitement}, and multimodal excitement features~\cite{merler2019highlights}. More recently, perception models have been trained to perform action spotting over large annotated datasets~\cite{giancola2018soccernet,deliege2021soccernetv2} to identify key moments in a match. Apostolidis et al.~\cite{apostolidis2021survey} survey the wider summarization literature. 
However, the highlight these approaches return is a ranked set of moments rather than a composed sequence, limiting viewer interpretability and control.

Language models have recently been applied to the same task. Structured representations such as scene graphs improve video understanding~\cite{yang2023pvsg,huang2025mindpalace} and reduce hallucination~\cite{yin2024woodpecker}, though in these systems the structure remains internal to the model. Our work extends this line by defining an action-centric schema consumed by both the generative pipeline and the viewer.

\subsection{Interfaces for casual game exploration}

Visualization research makes video navigable by pairing it with structured metadata. Matejka et al.~\cite{matejka2014videolens} presented Video Lens, which indexes baseball footage by attributes such as pitch type and player so fans can filter moments and play them back. Pavel et al.~\cite{pavel2014videodigests,pavel2015sceneskim} structure lecture and film video through aligned transcripts, and Deng et al.~\cite{deng2021eventanchor}
 reduce the effort needed to annotate racket sports events for such structured event indices.

Sports visualization has produced interfaces for both analysts and fans. SoccerStories~\cite{perin2013soccerstories} supported visual analysis of soccer phases, and Stein et al.~\cite{stein2018pitch} combined movement data with video. Closer to casual viewing, Lin et al.~\cite{lin2022quest} augmented sports video with embedded visualizations, and Lee et al.~\cite{lee2025sportify} answered tactical questions with narratives and embedded visualization. These systems show that structured metadata makes video navigable and supports fans engage with game context interactively. The structures they expose, however, are built for human retrieval alone. Our work explores a shared structure for both generative pipeline and human interfaces.

\begin{figure*}[!ht]
    \centering
    \includegraphics[width=0.85\textwidth]{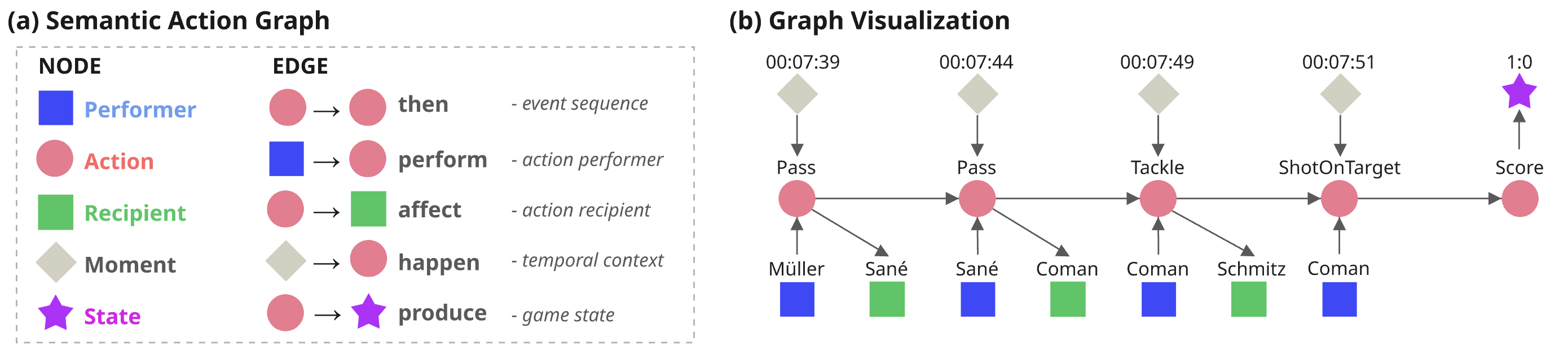}
    \caption{(a) Semantic Action Graph schema. Five node types record who acted, what occurred, who was affected, when it occurred, and what game state resulted; five edge types carry role, sequence, and outcome. (b) A goal sequence expressed in the schema. The \textbf{then} edge chains the action sequence, including two passes, the tackle, and the shot that \textbf{produced} the score.}
    \label{fig:graph}
    \vspace{-4mm}
\end{figure*}

%% file: docs/3_formative_study.tex
\section{Formative Study}
\label{sec:formative}

To understand how fans experience current highlights, we conducted 30-minute semi-structured interviews with six soccer fans (ages 18-55; 1 female, 5 male), including one casual viewer (F2), two regular fans (F1, F5), three die-hard fans (F3, F4, F6). Interviews covered viewing habits, highlight consumption, difficulties in finding specific content, and desired features. Recordings were transcribed and analyzed using reflexive thematic analysis~\cite{braun2006thematic}, with two researchers iteratively refining themes to consensus. 
Several gaps in current highlights impede fans’ engagement, which we categorized below.

\textbf{Content Gap.} Participants reported that highlights omit surrounding events and the build-up to key moments.
Four of six preferred seeing the actions leading to a goal over the finish alone: F3 wanted a clip to play \emph{``from when the action starts, the attack phase or previous defense phase, not just the actual moment of the goal.''}

\textbf{Narrative Gap.} Participants also wanted to understand why a moment mattered, not just what happened. There is often an issue with missing broader context and storytelling components: \emph{``Sometimes the highlight is just the moment of a goal, but I didn't know there was a red card before that''} (F1).

\textbf{Personalization Gap.} Current highlights use a one-size-fits-all approach that ignores individual preferences. Most fans we interviewed have clear preferences in highlight narrative types: \emph{``There is no ability to walk in onto a particular player. I’d love to be able to be more personalized''} (F4); \emph{``Everyone's different. People care about the big moments but I kind of want to see the build-up''} (F5).

\textbf{Discovery Gap.} Locating a specific moment was reported as difficult, particularly for less prominent events: \emph{``if it's a small moment, I usually have a hard time, like a penalty or a smaller moment that's not normally captured''} (F1).

These gaps translate into three requirements on the underlying representation, which we use as design goals below. \textbf{G1:} Play should be represented as connected sequences rather than isolated moments, so that build-up and consequence are available to both the system and the viewer. \textbf{G2:} The units the system reasons over should form a vocabulary viewers can also use to express preferences. \textbf{G3:} Individual moments should be addressable, so that a viewer can reach a specific event directly rather than by scrubbing.

%% file: docs/4_design.tex
\section{Semantic Action Graph}
\label{sec:sag}

\subsection{Schema}

A match is represented as a graph over five node types, as shown in Fig.~\ref{fig:graph}a. The \Identity{} \textbf{Performer} nodes record who carried out an action and the \Recipient{} \textbf{Recipient} nodes who was affected by it, both drawn from the match roster. The \Action{} \textbf{Action} nodes record what occurred, drawn from a sport-specific catalog (e.g., for soccer - Tackle, Foul, ShotOnTarget, Save, Score, YellowCard, and others). The \Moment{} \textbf{Moment} nodes carry temporal context in the form of game clock, video time, and frame index. The \State{} \textbf{State} nodes carry the game state an action produced, such as the score.

Five edge types connect these nodes with semantic relationships. Three encode the roles within a single event: 
\textbf{perform} (\Identity{}→\Action{}) links a performer to an action, \textbf{affect} (\Action{}→\Recipient{}) links an action to its recipient, and \textbf{happen} (\Moment{}→\Action{}) links a moment to the action occurring at it. The remaining two carry structure across events: \textbf{produce} (\Action{}→\State{}) links an action to the state it results in, and \textbf{then} (\Action{}→\Action{}) chains actions that are consecutive within a possession.  Fig.~\ref{fig:graph}b shows a goal expressed in the schema, in which a pass, a second pass, a tackle, and a shot are linked by \textbf{then} edges, ending in a \textbf{produce} edge to the resulting score.

\subsection{Design Properties}
The schema demonstrates
three key properties informed by the design goals in Sec.~\ref{sec:formative}. 

\textbf{Connected event sequences (G1).} With explicit edge definition in the graph, i.e. \textbf{then} edges connecting discrete actions, a highlight can be defined as a traversal of the graph rather than as a fixed time window around a detected event. For example, a highlight generation pipeline can add build-ups to the goal event by chaining back to a possession boundary rather than determining how many seconds before a goal as a parameter. The same edge structure also gives the viewer a compact representation of what a clip contains and in what order, which aligns with how a human viewer describes the event sequences.

\textbf{A shared, closed vocabulary (G2).} 
In our schema, action nodes come from
a fixed catalog of types, and performer and recipient nodes from the match
roster. Because the entities and events are defined in advance, the pipeline's generation freedom lies in phrasing rather than in facts. The same catalog is also presented to
the viewer as the filter panel (Fig.~\ref{fig:teaser}a), so the system
and the user shared same descriptive units. This closure also bounds what a
viewer can ask for. A request outside the catalog, such as one about crowd
celebrations, has no corresponding node and cannot be satisfied without extending
the schema.
This property makes the boundary of the system's capability explicit while allowing structured extension, such as adding new action types (e.g., different foul types) or descriptors for players (e.g., appearance, speed).

\textbf{Frame-addressable moments (G3).} Each \textbf{Moment} node carries a frame index of the event from the source feed, providing clear clip boundary when composing a highlight. A highlight generation agent can directly reference the moment nodes from the desired action sequence for accurate timestamps, avoiding incorrect generated time codes. For the viewer, the same property lets the viewer navigate the video through meaningful structure,(Fig.~\ref{fig:teaser}b), linking generated units to the footage to verify the narrative or localize specific moment effectively.

Overall, the semantic action graph schema builds upon the common metadata that professional event feed carries (e.g., IPTC~\cite{iptc2024sportschema}) and  adds two structural relations that metadata alone does not express. 
First, the \textbf{then} and \textbf{produce} relations make explicit the temporal and causal relationships between distinct action records, i.e., one action follows from another, and that an action changes the game state.
Second, the sport-specific vocabulary is fixed in advance and can be retargeted — participants, actions, sequence, and outcome are shared across most sports, while event granularity and narrative style are not, so a long build-up chain in soccer and a single last-minute shot in basketball are both expressible. Our contribution is not a new data model but a minimal relational layer that connects atomic event records to narrative structure.

Additionally, the schema is intentionally small.
It does not represent spatial position, formation, or tactical intent, which matter for analysis but are not required to cover much of the descriptive content of sports commentary. The schema therefore serves as a base for agentic highlight generation from metadata that professional feeds already provide, and it can be extended when spatial data such as player location, speed, or formation are available.

%% file: docs/5_system.tex
\section{\graph{}: A Highlight Generation Pipeline}
\label{sec:system}

\begin{figure*}[!t]
    \centering
    \includegraphics[width=0.85\textwidth]{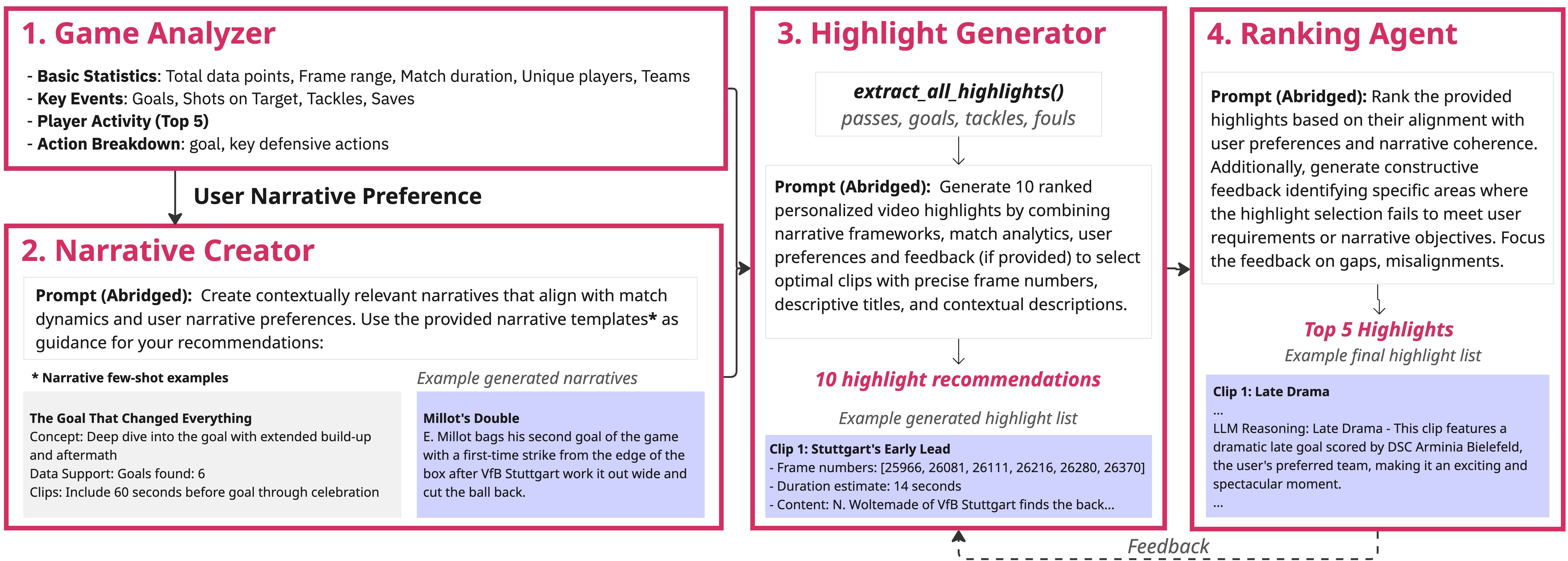}
    \vspace{-1mm}
    \caption{\graph{} Agentic Pipeline. All four modules operate on semantic action graph data; none reads video. 1. \texttt{Game Analyzer} computes match statistics. 2. \texttt{Narrative Creator} generates narratives tailored to user preferences. 3. \texttt{Highlight Generator} assembles 10 candidate clips bounded by \Moment{} \textbf{Moment} frame indices. 4. \texttt{Ranking Agent} selects the top 5 and provides feedback to (3). Modules 2-4 are LLM agents. }
    \label{fig:agents}
    \vspace{-4mm}
\end{figure*}

We built SportSAGE as a design probe to explore how a semantic action graph can support both agentic highlight generation and viewer-facing interpretation.

\subsection{Graph Construction and Agentic Pipeline}

We constructed the semantic action graph for a selected soccer match (i.e., 2024-2025 Pokal Cup Finale match between VfB Stuttgart and Arminia Bielefeld~\cite{espn}) from official league match event data. The raw event logs contain timestamped, frame-aligned records of actions, performers, and recipients at roughly 2-5 second granularity. Such format is routinely produced for professional soccer~\cite{bassek2025dataset}. We mapped the record onto the closed action catalog, including 13 actions covering possess, pass, tackle, foul types, shot types, kick types, and others. A 96-minute match yields 1,330 action nodes.

The SportSAGE highlight pipeline is composed of four modules that operate on these records, as shown in Fig.~\ref{fig:agents}. 

1) \texttt{Game Analyzer} is a deterministic component. It takes the graph as a CSV conforming to the schema, and computes match statistics and per-player activity profiles, removing arithmetic from the language model~\cite{schick2023toolformer}. 

2) \texttt{Narrative Creator} takes the computed statistics with a chosen narrative preference (action-focused, tactical analysis, player-centric, casual-viewing), and generates a highlight narrative using templates built around common soccer archetypes, such as \emph{Momentum Shifts}, \emph{Tactical Masterclass}, and \emph{Individual Brilliance}. 

3) \texttt{Highlight Generator} assembles ten candidate clips, each a set of nodes whose frame indices give the clip boundaries. 

4) \texttt{Ranking Agent} returns a top-five set and, through an optional iteration-limited reflection step~\cite{shinn2023reflexion}, feeds corrective feedback back. 

The three generative modules use Claude 3.5 Sonnet. 
Each receives the \texttt{Game Analyzer}'s summaries and a pool of candidate clips identified by their constituent records, without accessing the original video. 
The schema thus serves as the sole interface between the agent and the match. Statistics are computed deterministically, and clip boundaries derived from \textbf{Moment} frame indices rather than generated timecodes, ensuring the highlight output is factual.

\subsection{Interface}

The interface (Fig.~\ref{fig:teaser}) exposes the same schema through four components. A search panel filters by action type, player, and period, using the catalog of Sec.~\ref{sec:sag}. A node-link view renders the events in the current filter result or generated clip along a timeline, encoding team membership in node borders and player identity in thumbnails; selecting a node moves the video to that frame. A personalization panel offers four narrative focuses (\emph{Action}, \emph{Tactical}, \emph{Casual}, \emph{Player}) and team/player preferences. A highlight list presents each generated clip, and selecting a clip updates the node-link view to its constituent events. 
The frontend uses React with Cytoscape.js~\cite{franz2016cytoscape}.

%% file: docs/6_evaluation.tex
\section{User Feedback}

We collected feedback from 12 soccer fans (2 female, 10 male; ages 18-54; 6 casual, 3 regular, 3 die-hard) in a 40-minute online session on SportSAGE highlights and the graph utility. 
Participants first viewed both a default highlight set of the soccer match (five clips) and a customized highlight set with their preferred narrative focus (\emph{Tactical}, \emph{Casual}, or \emph{Player}). They provided feedback on their preference and perceived quality of the highlights.
They then used the SportSAGE interface to examine the AI highlight clips with semantic action graph and perform reasoning tasks on the interface, including locating specific game moments from graph, and linking the highlight events to its semantic structure. 

We summarized user feedback into two aspects: 1) AI highlight and narrative quality, and 2) graph layer usefulness.

\textbf{AI highlight and narrative quality.}
Participants were satisfied with the highlight sets produced by the agentic pipeline. On a 7-point scale (1 = poor, 7 = excellent), they rated overall highlight quality positively for both the default set ($M=5.25$) and the set matched to their chosen narrative focus ($M=5.25$). Perceived quality was therefore unchanged by personalization. The difference appeared in preference, where participants favored the customized narrative over the default (M = 5.92). This is consistent with the narrative and personalization needs found in the formative study. They praised clip coverage and accuracy (\textit{``it showed enough of the buildup to be engaging, the pivotal moment and some of the celebration''} (P4)) and valued the accompanying narrative for letting them \textit{``follow the progression of events easily''} (P7). Distinct narrative focus mattered for both die-hard and casual fan: 
\textit{``This [tactical] type of description is more favorable for me when I want to search and watch highlights or game re-cap''} (P8) and \textit{``It makes it easy to watch for a casual viewer like me who isn't too invested in soccer''} (P2). 
%

\textbf{Graph layer usefulness.}
Participants interpreted the semantic action graph correctly and used it to check narratives against the video and to recover game context. They rated the interface useful (M = 5.96) and completed the three tasks reliably, including filtering shot events (12/12), locating a specific player action (11/12), and interpreting video events from the graph (11/12).
In particular, participants valued the customizable nature and the granularity of the graph, which support fans \textit{``breaking down highlights to understand the plays better''} (P2). 

These preliminary insights suggest that AI highlights grounded in the semantic action graph yield satisfying quality and have the potential to address the content, narrative, personalization, and discovery gaps found in our formative study.

%% file: docs/7_discussion.tex
\section{Conclusion \& Future Work}
\label{sec:discussion}


We introduced the semantic action graph, a lightweight schema that represents matches as categorized, temporally ordered event sequences shared by both the agentic pipeline and the viewer-facing interface, making generative output in a narrative-rich domain more steerable and navigable.
SportSAGE applies this schema to soccer highlight generation, enabling personalized narratives that viewers can verify and navigate through the graph interface.

As a design probe, SportSAGE was evaluated on a single match with one model and no unstructured baseline, so we cannot yet attribute participants' satisfaction to the shared representation itself. Future work should compare against an unstructured baseline, extend the schema with spatial and tactical attributes, and explore constructing the graph directly from video rather than official feeds.